\documentclass[aps,twocolumn,prl,amsmath,superscriptaddress,floatfix,showpacs,showkeys]{revtex4-1}
\usepackage{graphicx}
\usepackage{hyperref}
\usepackage{color}
\usepackage{MnSymbol,wasysym}
\usepackage[normalem]{ulem}
\begin{document}

\title{Observation of strongly heterogeneous dynamics at the depinning transition in a colloidal glass}



\author{Nesrin \c{S}enbil}\thanks{these authors contributed equally}
\affiliation{Department of Physics, University of Fribourg, CH-1700 Fribourg, Switzerland}

\author{Markus Gruber}\thanks{these authors contributed equally}
\affiliation{Fachbereich Physik, Universit\"at Konstanz, 78457 Konstanz, Germany}

\author{Chi Zhang}
\affiliation{Department of Physics, University of Fribourg, CH-1700 Fribourg, Switzerland}

\author{Matthias Fuchs}
\affiliation{Fachbereich Physik, Universit\"at Konstanz, 78457 Konstanz, Germany}

\author{Frank Scheffold}
\affiliation{Department of Physics, University of Fribourg, CH-1700 Fribourg, Switzerland}

\begin{abstract}
We study experimentally the origin of heterogeneous dynamics in strongly driven glass-forming systems. Thereto, we apply a well defined force with a laser line trap on individual colloidal polystyrene probe particles seeded in an emulsion glass composed of droplets of the same size. Fluid and glass states can be probed. We monitor the trajectories of the probe and measure displacements and their 
distributions. Our experiments reveal intermittent dynamics
around a depinning transition at a threshold force. For smaller forces, linear response connects mean displacement and quiescent mean squared displacement. Mode coupling theory calculations rationalize the observations.
\end{abstract}
\maketitle
Tracking the passive or driven motion of a colloidal probe particle immersed in a complex environment, known as \emph{microrheology}, provides unique insights into the local mechanical and transport properties of materials \cite{waigh2005microrheology,squires2010fluid,furstMR}. Individual probe trajectories can be recorded and the heterogeneity of the dynamics can be studied directly \cite{furstMR,mason1997particle,manley2008high}.
In \emph{active} microrheology the motion of driven tracer particles is analyzed to probe the systems dynamics \cite{furstMR}.  Yet, it is often unclear if and when the probe faithfully samples the intrinsic thermal motion, especially at strong driving. Experimental probe trajectories, e.g.~in living cells \cite{Wang2013}, often shown deviations from classical drift-diffusive motion \cite{Hoefling2013}. 
\newline \indent Earlier experimental work explored the linear and nonlinear regimes in colloidal model systems \cite{habdas2004forced,wilson2009passive}. In their seminal work, Habdas et al. studied the nonlinear force to average velocity relations \cite{habdas2004forced}. Yet, experimental studies of the predicted, highly anomalous, spatio-temporal distributions of probe displacements %
are still lacking. Computer simulations suggest that probability distribution functions of the probe displacements in glassy systems are anomalously broad \cite{Winter2012,Reichhardt2010,gruber2016active,williams2006,winter2013jcp}, generally non-Gaussian, and often bimodal. The existence of two subpopulations of probes, one of which remains stuck in the glassy surroundings for long times, while the other moves (far) in the direction of the force, has been discovered in simulations of supercooled liquids \cite{williams2006,winter2013jcp} and of active particle systems \cite{Reichhardt2015}. Bimodality has been observed for short times in the motion of colloids in corrugated tracks \cite{Lee2006}, while power-law distributions are observed in granular systems close to jamming \cite{Chandelier2009,Reichhardt2010}. Hydrodynamic models \cite{koch_brady_1988}, mesoscopic models of glassy dynamics, like trap  \cite{Bouchaud1990} and continuous time random walk models \cite{Jack2008,Schroer2013,Gradenigo2016}, and  lattice models of transport in random media \cite{Benichou2013,Leitmann2017} have shown that the splitting into two populations lies at the origin of the intermittency in the probe motion.  Microscopic mode coupling theory has identified a threshold force for the delocalization (depinning) of a probe particle in glass \cite{gazuz2009active,gruber2016active}; average motion only sets in for forces larger than the threshold.  
\newline \indent Here we show experimentally and theoretically that structurally homogeneous colloidal suspensions around the glass transition exhibit heterogeneous and intermittent dynamics when a particle  is driven by an external force. Our experiments are performed on a near-ideal model system for hard spheres, which displays only weak dynamic heterogeneities in the quiescent state \cite{zhang2015structure,zhang2016dynamical}. Beyond a threshold force we observe strongly intermittent dynamics and bimodal van Hove distribution functions. For smaller forces, linear response connects the particle mean displacement and quiescent mean squared displacement. Using results from mode coupling theory we can rationalize the observations. Our findings highlight the important differences between quiescent and driven motion in crowded environments.
\newline \indent \emph{Experiment.--} 
  \begin{figure}[hbt]
 \centering \includegraphics[width=\columnwidth]{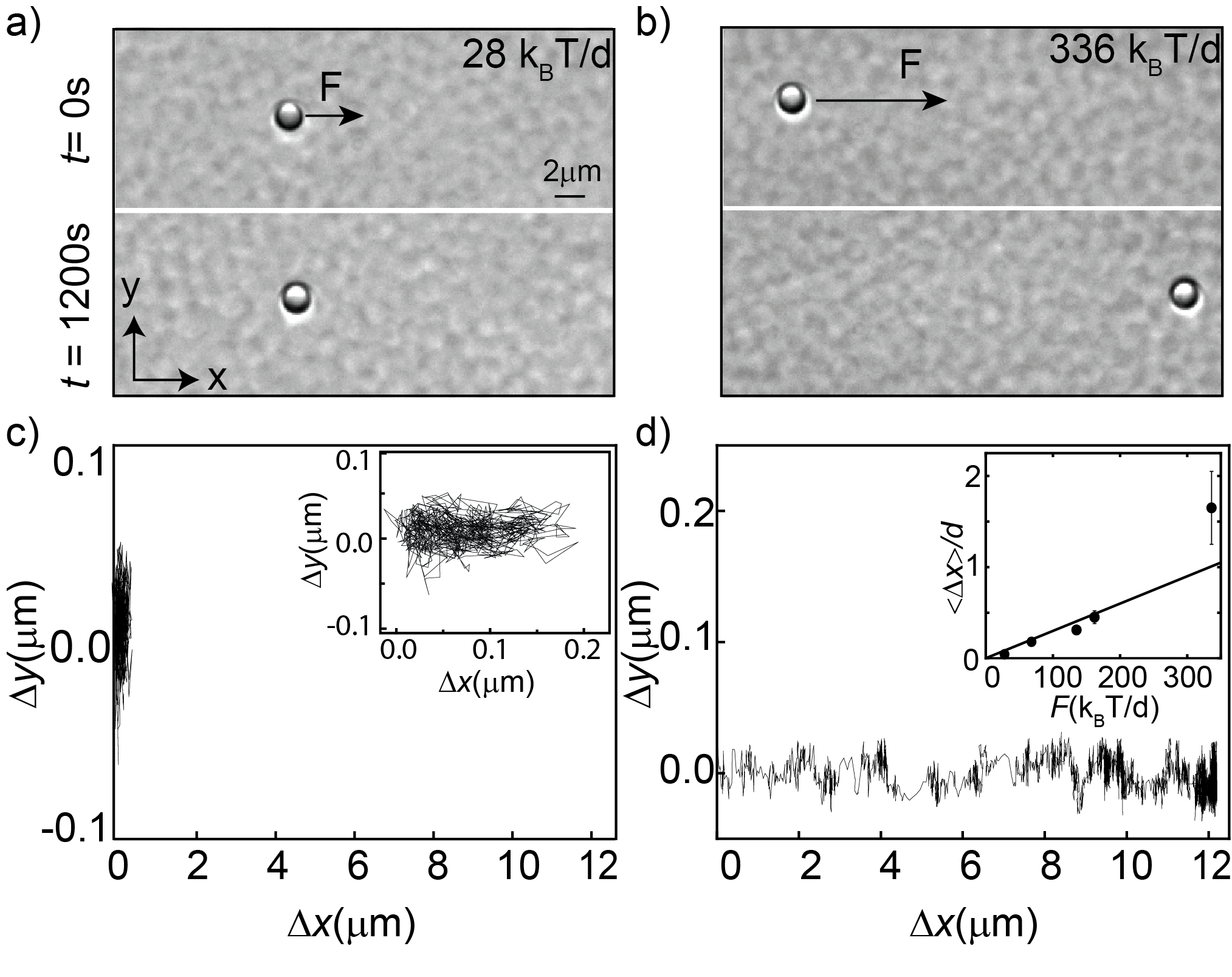}
\caption{Motion of a polystyrene probe particle seeded in a glassy emulsion at $\phi=0.601$. The diameter of the probe particle and the emulsion droplets $d=2 \mu$m are the same. The optical force is applied in x-direction and the results for two different laser power settings are shown. a) F=28 $k_B T/d$ and b) F=336 $k_B T/d$ with $k_B T/d=2.03$ fN. 
Probe position shown right before (top) and 20 min (bottom) after the constant force line trap has been activated. The lower panels (c,d) show a map of x-y positions of the probe particle over the whole duration of the experiment. Corresponding movies (accelerated 10$\times$) are included in the Supplemental Material \cite{SI}. Inset c): Enlarged view of the probe particle trajectory. Inset d): Average probe displacement at 60 seconds at each force. Solid line shows the linear response law Eq.~\eqref{linear_response} using the measured MSD at 60 seconds times the applied forces.}\label{fig2}
\end{figure}
We study experimentally the active microrheology of uniform oil-in-water emulsion droplets, mean diameter $d= 2.01 \mu$m, that show nearly hard-sphere behavior with an experimentally confirmed glass and a jamming transition at volume fractions of $\phi \simeq 0.59$ and $0.64$ \cite{zhang2016dynamical}, respectively. For the volume fractions considered, $0.53 <\phi < 0.61$, the hard-sphere like droplets are far from touching and there is no stress bearing network of contact points 
as is present in jammed emulsions \cite{liu2010jamming,jorjadze2013microscopic}.
The solvent and the emulsion droplets are refractive index and buoyancy matched and a small amount of added polystyrene probe particles of the same size provide optical contrast for laser trapping, see also {section 'Sample perparation protocol' in} the Supplemental Material (SI) \cite{SI}.
For each packing fraction, the probe particle mean square displacement (MSD) is first monitored in the quiescent state, without applying any force for 1200 seconds.
We find the well known slowing down of the long-time diffusion approaching  the colloidal glass transition and the cageing of particles for $\phi$ above it \cite{Hunter2012}. The results are in quantitative agreement with previous experiments  on  similar systems  \cite{Megen1998,zhang2016dynamical} and with calculations from mode coupling theory (see Fig.~S2 in SI \cite{SI}). 
The latter comparison confirms the mapping of short time diffusion coefficient $D_0$, density and length scale between measurements and theory. We position the probe particle in a gradient intensity line trap such that a constant force is created along the scan direction, while in the two perpendicular directions the particle motion is strongly confined, Figure \ref{fig2} \cite{crocker1999entropic}. 
From reference measurements in a simple viscous liquid 
we find that the force 
is constant $\pm 2 \%$ over a range of 25$\mu$m, corresponding to more than 12 particle diameters. The magnitude of the force can be adjusted by tuning the power of the laser $P_l$ and the forces that can be generated are of the order of several hundred femtonewtons.  
In the experiment the probe particle is first captured at a depth $z = (2\sim6) d$  using a single-beam point trap. Larger depths are not accessible due to residual scattering and optical aberrations. The finite depth might induce a small numerical shift of the results due to wall effects, similar to the case of simple fluids \cite{sharma2010high}, but we do not expect significant qualitative changes of the dynamics. Subsequently we align the optical tweezer and the probe particle position and at $t=t_0$ switch the optical configuration to apply a constant force $\vec{F}(t>t_0)=F\vec{e}_x$ in the $x-$direction parallel to the surface of the sample. The image acquisition is started at a time $t$ with a delay of $t-t_0 = 0-0.2$s for the smaller and $t-t_0 \sim 0.5-1$s for higher forces, the latter due to an earlier realization of the experiment. The accuracy of tracking the probe particle is approximately $\pm 30$nm \cite{zhang2016dynamical}. The main uncertainty arises from the the unknown delay $t-t_0$. We take account of this by plotting a systematic error interval, as shown in Fig.~\ref{fig3} (there and in all following plots $t_0=0$ is set). Data points for larger displacements and longer times, which are our main interest, are not affected due to the logarithmic scales employed. 
Using a digital camera we record five images of the sample per second and subsequently track the position of the particle for each frame using standard procedures \cite{furstMR}. For each force the experiment is repeated more than ten times on different probe particles and in some cases up to forty times. 
\begin{figure}[ht]
 \includegraphics[width=\columnwidth]{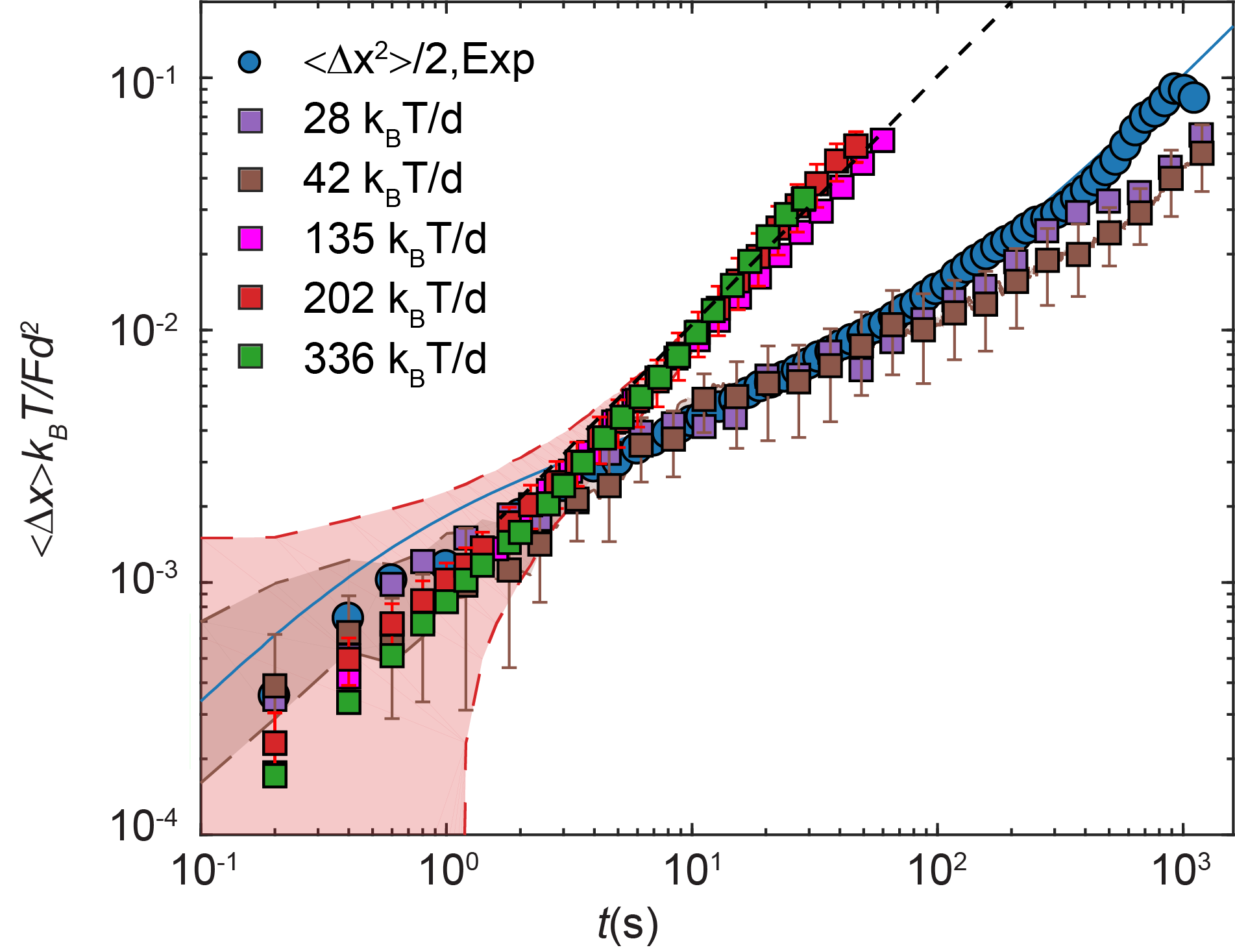} 
 \caption{\label{fig3} 
 Mean motion of the probe in a supercooled liquid: The rescaled mean displacement (MD) $\langle \Delta x(t) \rangle/F$ at  $\phi= 0.535$  are shown as squares for different forces as labeled in
$k_BT/d$. Overlap with the equilibrium MSD curve  (blue circles) and MCT prediction for the MSD (solid line) for the lower forces verifies the validity of linear response, Eq.~\eqref{linear_response}. 
For the higher forces the dashed line marks the linear drift regime with $\langle \Delta x(t) \rangle/(F d^2/k_BT) \simeq 10^{-3} t$/s. 
The shaded area marks the systematic error due to the uncertainty with respect to $t-t_0$ for the higher forces (red) and for lower forces (brown). The range of uncertainty is $[t,t+\delta t]$ and $[\Delta x(t), \Delta x(t+\delta t)]$ with $\delta t = 0.2 (1)$s for the lower (higher) forces. We estimate  $\Delta x(t+\delta t)$ using Eq. (\ref{linear_response}).  Error bars mark the statistical errors. } 
\end{figure}
\begin{figure*}[ht]
 \centering \includegraphics[width=2\columnwidth]{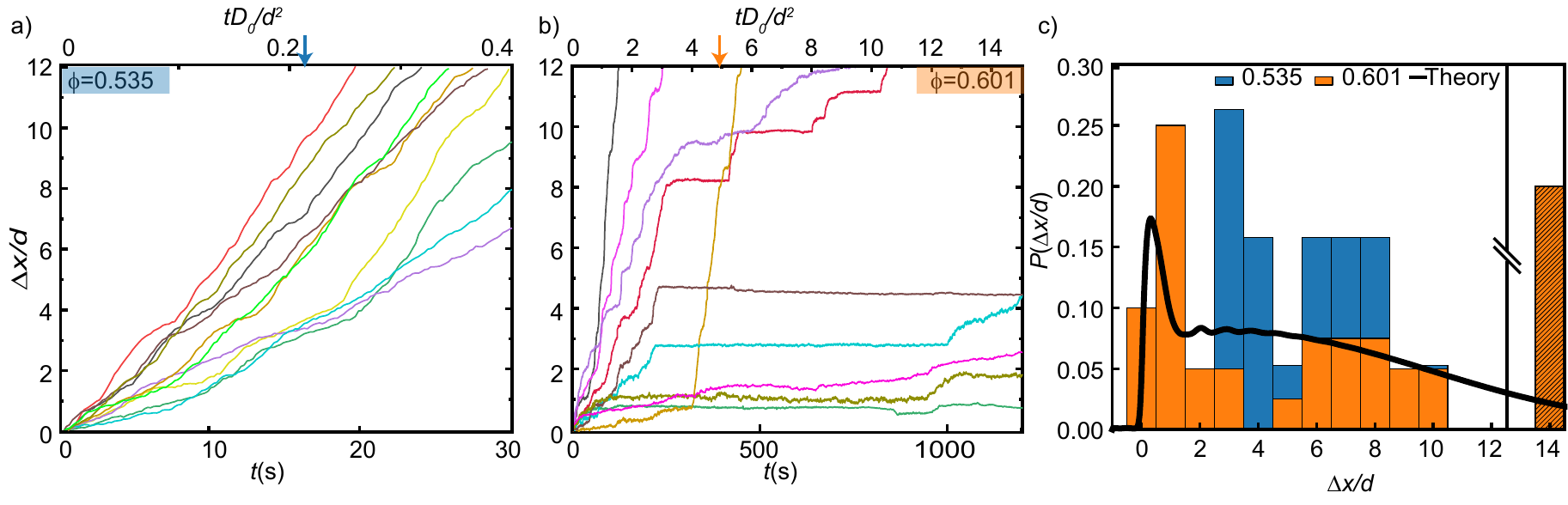}
\caption{
\label{fig5} Intermittent dynamics in the glass. Panels (a) and (b) show typical individual probe displacement curves in the direction of the applied force $F$ in the liquid (a, $\phi=0.535$, $Fd/k_B T = 336$) and the glass (b, $\phi=0.601$, $Fd/k_B T = 336$).  The time window in b) is enlarged $40 \times $ to cover the constrained probe displacements in the glass. Arrows indicate the times, at which the histograms in panel (c) were taken. Times were chosen such that the median of the data is the same ($\widetilde{\Delta x}=5.7d$). Particles, which have reached the end of the trap are collected in the bin at 14. To illustrate the critical behavior, we show a theory curve (solid line) at the threshold force $F_c=34.4 k_B T / d$ for a time, where the median is similar. 
}\end{figure*}
\newline \indent \emph{Linear response and intermittent dynamics in the fluid state.--}
The motion of the probe particle subject to the external force will depend on the strength of the forces and the emulsion concentration. We carefully analyze the particle trajectories for
two compositions, one in the viscoelastic fluid regime ($\phi=0.535$) and one in the glass  ($\phi=0.601$).
When applying a relatively small force  the mean displacement (MD) of the probe should obey the linear response relation \cite{Risken}:
\begin{equation} \label{linear_response}
\langle \Delta x(t) \rangle = \frac{\langle \Delta x^2(t)\rangle_{\rm eq}}{2 k_BT} F\;,
\end{equation} 
which identifies the equilibrium 1D-MSD $\langle \Delta x^2(t)\rangle_{\rm eq}$ divided by $2k_BT$ as time-integrated mobility. Equation~\eqref{linear_response} predicts that the ratio $\langle \Delta x(t) \rangle / F$ collapses onto the MSD (in units of $2 k_B T$) for times and forces where nonlinear effects are negligible. Interestingly, to our best knowledge, this law has never been tested experimentally for strongly correlated colloidal liquids. 
Figure \ref{fig3} shows that for lower forces the linear response relation holds in the supercooled state for a wide window in time where the probe explores the glassy cage and its slow relaxation. 
Increasing the laser power and employing forces of order $100 k_BT/d$, the measured MD 
speed-up at long times and approach a linear drift, Fig. \ref{fig3}.
The force-induced escape from cages dominates relative to the one by thermal fluctuations in the viscoelastic fluid state. MCT supports these conclusions, see Fig.~S5 in SI \cite{SI}; quantitative differences exist in the magnitude of the effects.  
\newline \indent \emph{Depinning and intermittent dynamics in the glass.--}  Observation of linear response in the glass is challenging, because the displacements are small. While the uncertainty in establishing the starting point of the trajectory affects the MD data for short times, we still find linear response for $t=60$s as shown in the inset in panel d) of Fig.~\ref{fig2}. The line shows the prediction based on the measured force-free MSD at 60s, which is long enough to not suffer from the short time uncertainties and short enough to avoid problems due to a possible drift of the system. The data for all times with the full uncertainty analysis is shown in Fig.~S7 in the SI \cite{SI} and confirms that any time between 50s and 200s would give the same results.
MCT predicts a threshold force of $F_c d/k_B T=34.4$ in the glass \cite{gazuz2009active,gruber2016active}, which is well within the experimentally accessible range. It should be noted that this transition is quite sharp in theory, i.e. a small variation in the force causes a large variation in the behavior of the mean displacement and thus a phase diagram separating delocalized and localized regimes can be established as shown in \cite{gruber2016active}. In the experiments and previous simulations \cite{gruber2016active} this phenomenon appears over a broader range of forces. This makes it more difficult to find the threshold force in the experiment. From our data, inset Figure \ref{fig2} d), we estimate it to be $F_c d/k_B T \sim 135-300$, which is larger than the MCT prediction, and slightly larger than the simulation results \cite{gruber2016active}. 
\begin{figure}[ht]
 \centering 
 \includegraphics[width=.97\columnwidth]{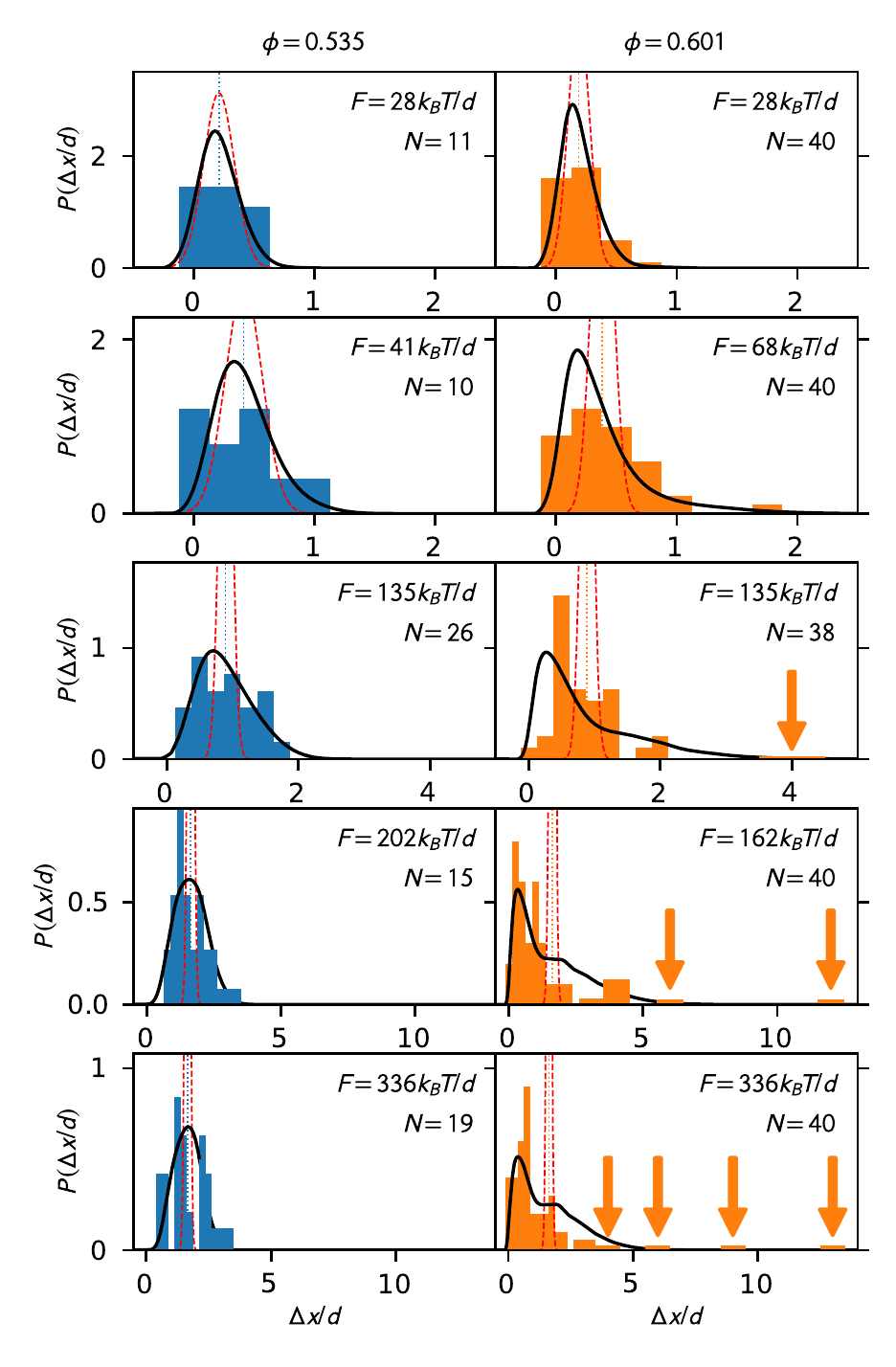}
\caption{\label{fig4} Strong forces induce intermittent displacements.  We compare experiments (histograms, $N$ experiments for $\phi=0.535$ and $\phi=0.601$) and theory (solid lines) for different applied forces at times, where the mean displacement (dotted vertical line) is the same. The left panels show the van-Hove function in the liquid $(\phi=0.535$), while the right panels show it in the glass ($\phi=0.601$) for forces increasing from top to bottom. The displacements are determined from the largest displacement available in the glass: $\langle \Delta x \rangle = 0.2 d, 0.4d, 0.9d, 1.6d, 1.6d$ (from top to bottom).  Arrows indicate single observations. The forces in theory  are chosen such that the similarity (as introduced in \cite{froufe-perez2017band}) between the histogram and PDF is maximized. For comparison, we plot a Gaussian (dashed red line) with the same mean and variance given by the quiescent MSD at the same time.
The times at which the histograms are taken are for $\phi=0.535$: 32.2s,
64.8s, 7.6s, 9.0s, 5.8s and for $\phi=0.601$: 973s, 801s, 644s, 642s, 
59s (from top to bottom, i.e.\ low force to high forces).
}\end{figure}
\newline \indent  We now turn our attention 
to the dynamics at the depinning transition.
In Figure \ref{fig5}, we show several trajectories $\Delta x(t)$ of the probe particle for an emulsion volume fraction of  $\phi=0.601$ (glass) at a force $F\gtrsim F_c$  close or slightly above the depinning transition. For comparison, we include trajectories in the fluid ($\phi=0.535$) for a similar force. Also the complete histograms for large median displacements $\langle\widetilde{\Delta x}\rangle=5.7d$ are compared for fluid and glass sample. This value is determined by the largest median displacement measured in the glass. Clearly, the motion is far more intermittent in the glass than in the fluid state and the displacement distribution  is far broader. The probability distribution function (PDF) of displacements in force direction, viz.~the van Hove function $G^s(\Delta x,t)$ 
\cite{furstMR} can also be calculated from theory. MCT predicts bimodal shapes of pinned and mobilized sub-populations close to the depinning force $F_c$. A PDF at $F_c$ and identical median displacement is added for comparison in Fig.~\ref{fig5}(c). It correlates well with the sampled histograms. 
\newline 
To answer the question whether the force-induced motion differs  qualitatively from the (intrinsic) thermally induced particle motion, we compare the PDF of displacements for a state where thermal motion is active ($\phi=0.535$) and one where it is not ($\phi=0.601$).
Figure~\ref{fig4} shows histograms of the PDF at fixed average displacement $\langle \Delta x\rangle$ comparing data at the same or comparable force settings from fluid (left) and glass (right panels) samples.   The chosen distances from $\langle \Delta x\rangle=0.2d$ (upper) to $1.6d$ (lower panels) correspond to the largest mean displacements available in this setup. The experiment ends when the first particle reaches the end of the line trap or after around 1000s. We compare forces below and above the depinning transition for the glass and choose similar forces for the liquid. 
To illustrate the non-Gaussian behavior, a Gaussian distribution with the same average displacement (viz.~MD from Fig.~\ref{fig3}) and quiescent variance (viz.~MSD at the same time)  is compared to the data. Also MCT calculations for the same MD values are included. Because of the force mismatch in the theory, MCT-forces are fitted  to the histograms optimizing the similarity (over a range of mean displacements) following ref.~\cite{froufe-perez2017band}. In the fluid state for the lower force, the PDF of the probe still  resembles the Gaussian solution of the drift-diffusion equation expected in dilute systems \cite{furstMR}. In the glass at this force, however, the PDF extends to larger displacements than the shifted Gaussian even though it has the same average  $\langle \Delta x\rangle=0.4d$. The differences between the fluid and glass PDF become larger with stronger forces. In glass where force induced motion dominates,
some probes remain localized within their cages, while some other probes can escape their neighborhood and reach displacements comparable to the bath particle size or larger. This reveals the heterogeneity in the cage strength and the collective origin of the force pinning the probe.  MCT predicts the appearance of an exponential tail when approaching $F_c$ \cite{gruber2016active}  which is compatible with the data-histogram albeit not clearly resolved due to the limited number of experiments $N$.  Even stronger heterogeneity in the probe motion is visible at the largest forces.
 The interpretation suggested by theory and simulation \cite{williams2006} is that the PDF develops a bimodal shape in the glass consisting of one sub-population of pinned and another sub-population of mobilized particles.  In the fluid state, the additional bath motion narrows the PDF as cages open more uniformly  by thermal fluctuations.  Bimodal PDF arise in the MCT calculations in a range of forces below and close to the glass transition (not shown) which implies that a characteristic force remains meaningful also in fluid states; it separates intrinsic from force-induced cage breaking processes.
\newline \indent \emph{Discussion and conclusion.--} In summary, we have shown that  force-induced intermittent motion can be observed and quantified in glass-forming dispersions, tracking colloidal probes manipulated in an optical line-trap.  Linear response rationalizes the behavior for small forces of the order of ${\cal O}(10 k_BT/d)$ for a broad time window. Force-dominated motion sets in at longer times, including in glass states where a force threshold $F_c$ needs to be overcome. Depinning and cage-breaking is characterized by intermittent probe motion and anomalous broadening of the displacement probability distribution. 
Theory rationalizes the observations and predicts bimodal distributions, where a sub-population of particles remains trapped while another subpopulation moves far. Intermittent motion arises in undercooled fluid states and gets stronger when approaching the glass transition, as correlates with the growth of dynamically  heterogeneous regions  seen in quiescent dispersions \cite{weeks2000three}. Yet, it is strongest in glass  where only smaller cooperative clusters were observed without force. This indicates that the link between intermittent motion in active microrheology  and dynamical heterogeneities is more indirect than previously discussed \cite{Gradenigo2016,Jack2008}.  A qualitative comparison with mode coupling theory is possible. In the experiment, anomalous dynamics is observed over a broader range of forces than predicted theoretically. 
\begin{acknowledgments}N.S., C.Z. and F.S. acknowledge funding by the Swiss National Science Foundation through project 169074 and through the National Center of Competence in Research \emph{Bio-Inspired Materials}. M.G. and M.F. acknowledge financial support from the Deutsche Forschungsgemeinschaft in Project P3 of Grant FOR1394. We would like to thank Veronique Trappe and Thomas G. Mason for discussions.\end{acknowledgments}

%

\clearpage
\section*{Supporting Information}
\section*{\c{S}enbil et. al.} 
\vspace{2mm}

\setcounter{equation}{0}
\setcounter{figure}{0}
\setcounter{table}{0}
\setcounter{page}{1}
\makeatletter
\renewcommand{\theequation}{S\arabic{equation}}
\renewcommand{\thefigure}{S\arabic{figure}}
{\bf{Sample preparation protocol.--}}
We prepare uniform oil-in-water emulsions
as described previously \cite{zhang2015structure,zhang2016dynamical}. The oil phase consists of PMHS (Poly(methylhydrosiloxane), Sigma-Aldrich cat. no:176206). First, a crude emulsion is prepared using a custom made Couette shear cell, stabilized with SDS (Sodium dodecyl sulfate) and subsequently size segregated by depletion fractionation until the desired size polydispersity of approximately 12\% (standard deviation/mean) is reached \cite{mason1996monodisperse}. The droplet mean diameter of $d=2.01  \pm 0.05$ $\mu$m and the polydispersity were determined by multi-angle dynamic light scattering (LS Spectrometer, LS Instruments, Switzerland). Confocal microscopy of a dye-labeled sample were found in agreement with these results.  We replace SDS by the block-copolymer surfactant Pluronic F108 (BASF, Germany) to achieve steric stabilization of the emulsion droplets. For the active microrheology experiments, the solvent and the emulsion droplets are refractive index and buoyancy matched at room temperature $T=22^\circ$C by replacing the aqueous solvent with a mixture of water, DMAC and Formamide, volume ratio 6:4:1 \cite{zhang2015structure}. As shown previously we can control the droplet volume fraction to better than $3\times 10^{-3}$ when taking the jamming condition at $\phi_J=0.642$ as a reference \cite{zhang2015structure,zhang2016dynamical} (see Figure \ref{figsmDLSrot}). In the present study we add a small amount of polystyrene particles (PS), volume fraction about $10^{-4}$, of the same size to serve as microrheological probes. The PS particles (micromod, Germany, product 01-54-203) have a diameter $2 \mu$m  and polydispersity 5\% (standard deviation/mean, supplier information). The particle are stabilized with a coating of polyethylene glycol (mol. weight 300 g/mol). The refractive index is $n\simeq 1.6$, significantly higher than the solvent and the droplets $n=1.40$ which provides contrast for the optical tweezer and allows us to apply a force on the particle.  The sample is filled in a custom made plastic cell (thickness 200 $\mu$m, area $3.1$mm$^2$), and closed with glass cover slip placed under a microscope objective combined with a laser tweezer setup.
\newline We centrifuge the fractionated emulsion at 4$^\circ$C and 4000rpm overnight. Lowering the temperature to 4$^\circ$C induces a slight density mismatch between the emulsion droplets and the solvent and consequently a solid plug is formed at the bottom of the cuvette. Then, trace amounts of polystyrene probe particles are dispersed into the sample. By choosing a sufficiently high starting concentration for the emulsion we make sure that the PS particles do not sediment or cream and remain dilute throughout the sample, even though they are not perfectly buoyancy matched with the bath (The density of the solvent and the oil is 1.006 g/ml at 22$^\circ$C while the density of the polystyrene particles is 1.03 g/ml). Samples with different compositions are obtained by dilution with the solvent phase. 
\smallskip
\newline \indent {\bf{Determination of the unjamming transition by dynamic light scattering.--}}
The accuracy of directly measuring droplet fractions by drying and weighing is limited, as discussed e.g. in \cite{royall2013search}.  Therefore we chose dynamic light scattering (DLS) as a sensitive tool to  determine the unjamming point of our emulsion precisely which we can then use as a reference.  
\begin{figure}[h]
\includegraphics[width=.85\columnwidth]{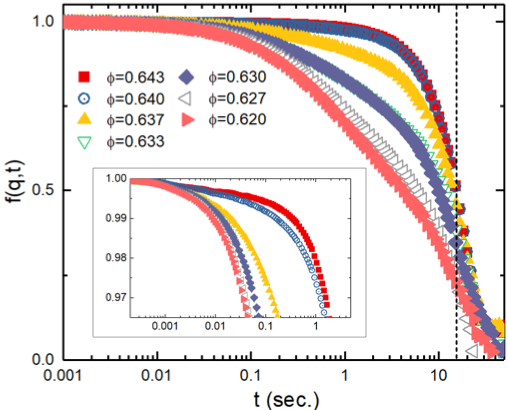}
\caption{\label{figsmDLSrot} Intermediate scattering function for different emulsion concentrations $\phi$. Measurements are taken at an angle of 90$^\circ$ which corresponds to $q · d$ = 37.8 for the incident laser light of wavelength $\lambda$ = 660nm. Inset: Enlarged view of the same data for $f(q,t)$ close to one. }
\end{figure}
The sample at this composition is then set to $\phi_J = 0.642$ as predicted by computer simulations for hard-sphere or nearly hard spheres systems with 12\% polydispersity \cite{desmond2014influence,zhang2015structure}.  Dilution of the stock emulsion can be done very precisely since we can measure the weight and volume of the added solvent accurately and we know that the density of the solvent and the oil are the same.  To determine the unjamming point we dilute an as-prepared jammed stock emulsion in small steps and record the normalized intensity-intensity autocorrelation function $g_2(t)-1$ using a commercial light scattering goniometer (LS Spectrometer, LS Instruments, Switzerland). All measurements are taken at an angle of $90^\circ$ which corresponds to $q \cdot d=37.8$ for the incident laser light with wavelength $\lambda=660$nm.  Here we take advantage of the weak but finite scattering contrast after (near) refractive index matching with the solvent as well as the presence of tracer particles. To obtain proper ensemble averages, the sample is put on a rotation stage and rotated very slowly at approximately 0.001 rpm \cite{xue1992nonergodicity}. In the jammed state, the elastic modulus of the sample is rather high and droplets are in contact $f(q,t)_{sample}\simeq 1$. The rotation, however, induces a terminal decay of $f(q,t)$ at about $t=10-20$ sec as can be seen clearly in Figure \ref{figsmDLSrot}. Upon dilution of the sample, the free volume per particle becomes finite and we observe an additional decay, associated with the local motion of the droplets (also known as the $\beta$-relaxation). Figure \ref{figsmDLSrot} clearly shows the sensitivity of our experiment to this effect. Initially the curves are nearly flat $f(q,t)\simeq 1$ indicating a highly elastic solid which is followed by the decay due to the rotation of the cuvette.  At some point we observe first deviations (inset Figure \ref{figsmDLSrot}). Further decreasing the concentration by only $3\times 10^{-3}$ the differences become dramatic. This shows that our DLS-experiments are very sensitive to the unjamming transition. DLS allows us to determine the jamming/unjamming transition with an accuracy better than $\Delta \phi \sim 3\times 10^{-3}$. 
\smallskip
\newline \indent {\bf{Fitting of time and length scales.--}}
To compare experimental results and MCT predictions, we need to find the proper packing-fractions, time- and lengthscales. The corresponding packing-fraction is not fitted but calculated by adjusting the glass transition point via $\phi_{MCT}/0.516 = \phi_{exp}/0.59$. The lengthscale in the system is determined by the bath particle diameter $d$. The MCT timescale $\bar{t}$ is set by the short time diffusion coefficient via $\bar{t}=d^2/D_0$, which coincides with the bare diffusion coefficient in dilute samples. For dense suspensions, short time diffusion is slower due to interactions (steric and hydrodynamic) and therefore needs to be measured or fitted. Since the MCT glass transition occurs at lower packing fractions, the localization length in the glass is larger than in the experiment. Therefore, we also allow for a fitting of the length-scale $\bar{d}$, as has been done in earlier comparisons \cite{sperl2005nearly-logarithmic}. We do a least squares fit in the log-log-representation to determine the time-scale and the length-scale for each packing fraction individually. Experimental data and fits are shown in Fig. \ref{fig1}. The fit-parameters can be found in Table \ref{fit_scales}. These length- and time-scales are then used for the following comparison. 
\begin{figure}
 \centering \includegraphics[width=1\columnwidth]{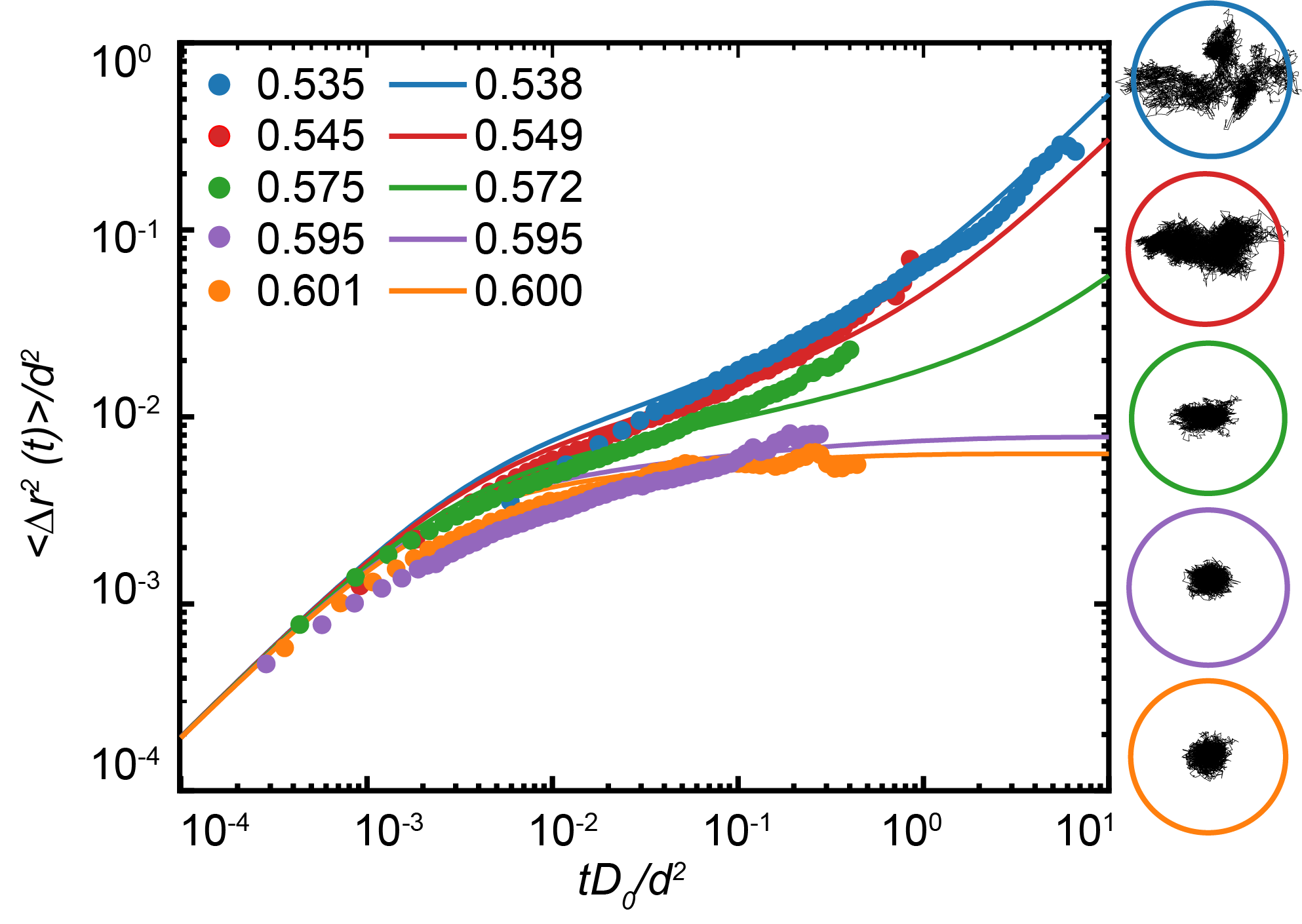}
\caption{Thermally driven motion ($F\equiv 0)$. 2D mean square displacement (MSD) 
$\langle \Delta x^2(t)+\Delta y^2(t)\rangle$ of the emulsion droplets at different packing fractions $\phi$ (filled circles). Solid lines are calculated MSD obtained from MCT.   
For $\phi \ge \phi_g\simeq 0.59$ the sample is dynamically arrested. 
Right column: Circles show corresponding examples for the displacements ($\Delta$x(t), $\Delta$y(t)) tracked in a 2D-plane for same $\phi$ over a duration of 1200 seconds. The diameter of the circles is d=$ 2\mu$m.}\label{fig1}
\end{figure} 
\begin{table}[h]
\caption{\label{fit_scales}Corresponding parameters between experiments and MCT}
\begin{tabular}{ccccc}
$\phi_\text{exp}$ & $\phi_\text{MCT}$ & $\bar{t}$ (s) & $D_0 = d^2/\bar{t}$ ($\mu$m$^2$/s)& $\bar{d}/d$ \\\hline
0.535 & 0.47 & 167 & 0.0237 & 0.795 \\
0.545 & 0.48 & 1090 & 0.00366 & 0.965 \\
0.575 & 0.50 & 2310 & 0.00173 & 1.21 \\
0.595 & 0.52 & 3500 & 0.00114 & 1.20 \\
0.601 & 0.525 & 2790 & 0.00144 & 1.12 \\
\end{tabular}
\end{table}
\smallskip
\newline \indent {\bf{Constant force calibration.--}}
While a conventional optical \emph{single-beam gradient force trap} is characterized by a roughly linear force-distance relation, in our experiments, we realize an $x-$position independent constant force by using a time shared line trap configuration (Figure \ref{FigTweezer}). We use a high power infrared laser $\lambda=1064$nm as a light source (IPG Photonics, USA, 10W). 
\begin{figure}[h]
\includegraphics[width=.9\columnwidth]{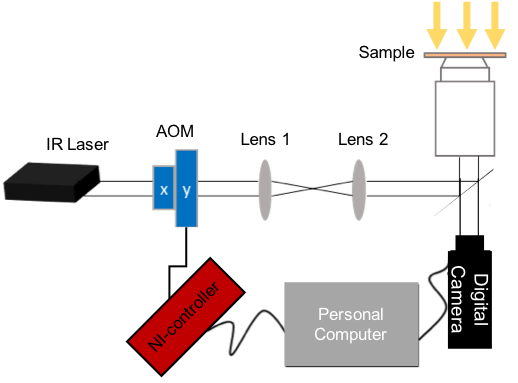}
\caption{\label{ltweezschem}Schematic of the optical tweezer and imaging setup. IR-laser: Infrared laser $\lambda$=1064nm with up to 10Watt output power. Two crossed acousto optical modulators (AOM's). $L_1$ and $L_2$: Lenses with focal length $L_1=100$mm and $L_2=250$mm. IR reflective dichroic mirror, inverted Nikon TS100 microscope body with a 60X/1.40 Nikon objective employed to form a tightly focused spot inside the sample. The AOM's and the back aperture of the objective are located in conjugate planes. The $z-$position can be adjusted with the TS100 body. On top of the objective a glass cell containing the sample is placed on a $x-y$ translation stage. White light illumination from the top and image recorded of the sample with a digital camera through a dichroic mirror. The control of the AOM and the data acquisition is performed with two National Instruments controller cards (NI-BNC-21110 and NI-PCI-6229) and a personal computer (PC).}\label{FigTweezer}
\end{figure}
To be able to compare different experiments we measure the laser power $P_l (\text{mW})$ after the AOM (before the lens 1) using a set of attenuators and a power meter (Thorlabs, USA). The actual laser power incident on the sample is about 30 $\times$ higher, typically of the order of $0.1-1$ Watt distributed over a line of length $x_0=40 \mu$m.. The laser beam is expanded to fill the back aperture of a microscope 60$\text{X}$ microscope objective. We can control the incident angle, perpendicular to the direction of propagation, using two fast acousto optical modulators (AOM's). The angular displacement allows us to control the $x,y$ position in a the focal plane of the objective, which is adjusted to be parallel to the glass interface of the sample cell layer. In the present experiment we only displace the beam along the $x-$ axis to create a line trap.
\begin{figure}[h]
\includegraphics[width=.85\columnwidth]{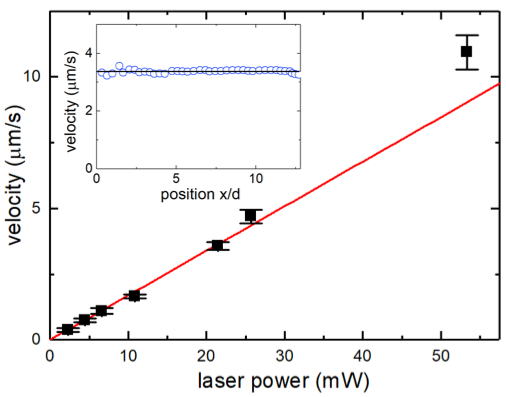}\label{ltweezcal} 
\caption{Velocity $v$ of a probe particle in the bulk of a simple liquid with the same refractive index as the emulsion sample. The probe particle velocity, and thus the applied force, increases proportionally with the incident laser power setting $P_l$. Inset: For a given laser power the velocity (shown: $v=3.37 \pm 0.06 \mu$m$/$s, $P_l$=21.5 mW) is constant over an $x-$ range of more than 12 particle diameters $d$. In the stationary state the applied force is balanced by the stokes drag $F_\text{Stokes}=\zeta v$. Using the bulk viscosity of the solvent $\eta \simeq$ 4 mPa s to calculate the Stokes drag $\zeta=3 \pi \eta d$ we find $F\simeq 12.8 \times P_l$ fN/mW.}
\label{ltweezcal}
\end{figure}
The voltage sequence was designed such that the OT is distributed along a line (e.g. $x \in [0, x_0]$) in a random manner. Moreover, the density probability function $\rho$ of the OT position is set to increase linearly $\rho(x) = kx$ with $\rho(0)=0$ and $k=$constant. The force (along the line) acting on the probe located at $x$ from the OT positioned at $x+\xi$ can be denoted as $f(\xi)$, where $\xi \in [-\xi_0, \xi_0]$. Here $\pm \xi_0$ denotes the finite range over which the OT can influence the probe and  $\xi$ is the relative distance between the OT and the probe. For positions far from both ends of the trap, i.e. $ x \in [\xi_0,x_0-\xi_0]$, the total force applied on the probe can be written as
\begin{equation}
\centering
F(x) = \int_{-\xi_0}^{\xi_0} \rho(x+\xi)f(\xi)d\xi.
\label{eq1}
\end{equation}
Since $\rho(x) = kx$, equation (\ref{eq1}) can be rewritten as
\begin{equation}
\centering
F(x) = \int_{-\xi_0}^{\xi_0} \rho(x)f(\xi)d\xi  +\int_{-\xi_0}^{\xi_0} \rho(\xi)f(\xi)d\xi.
\label{eq2rr}
\end{equation}
Since $f(\xi) = - f(-\xi)$, the first term of equation (\ref{eq2rr}) is zero, meaning that $F(x)$ is independent of $x$. Thus, the total force is constant $F(x)\equiv F$, within the range of $ x \in [\xi_0,x_0-\xi_0]$. Since roughly $\xi_0 \simeq 2 \lambda \simeq 2 \mu m$ \cite{woerdemann2012introduction} for objectives with numerical aperture $\simeq 1.4$ and in our case $x_0 \simeq$ 40 $\mu m$, we therefore obtain a constant force over a large $x$-range. We note that all components of the forces scale with the laser power selected. While $F_y$ and $F_z$ point towards $\Delta y, \Delta z=0$, $F_x\equiv F$ is constant and points in positive x-direction.
\newline To verify the successful implementation of the constant force line trap we measure the velocity of the PS probe particle
in a pure solvent mixture of water and DMAC (volume ratio 1:1.15) with a viscosity $\eta \simeq$ 4 mPa s and a refractive index matched to the index of the emulsion droplets \cite{aminabhavi1995density}. Figure \ref{ltweezcal} a) shows that the velocity is indeed constant, within better than 2\%, over a range of more than 12 particle diameters in $x-$direction corresponding to more than $25 \mu$m. Moreover, as shown in Figure \ref{ltweezcal},  the velocity increases linearly with the power $P_l\text{(mW)}$ of the laser. This demonstrates that we can precisely control the constant force applied to the probe particle. From the velocity measurements we can estimate that we are able to apply forces in the range of several hundreds of femtonewtons. In the emulsion, residual scattering from the droplets might somewhat perturb the line trap. This is discussed in the Section \emph{Linear response calibration. 
}

\begin{figure*}[ht]
 \includegraphics[width=.95\columnwidth]{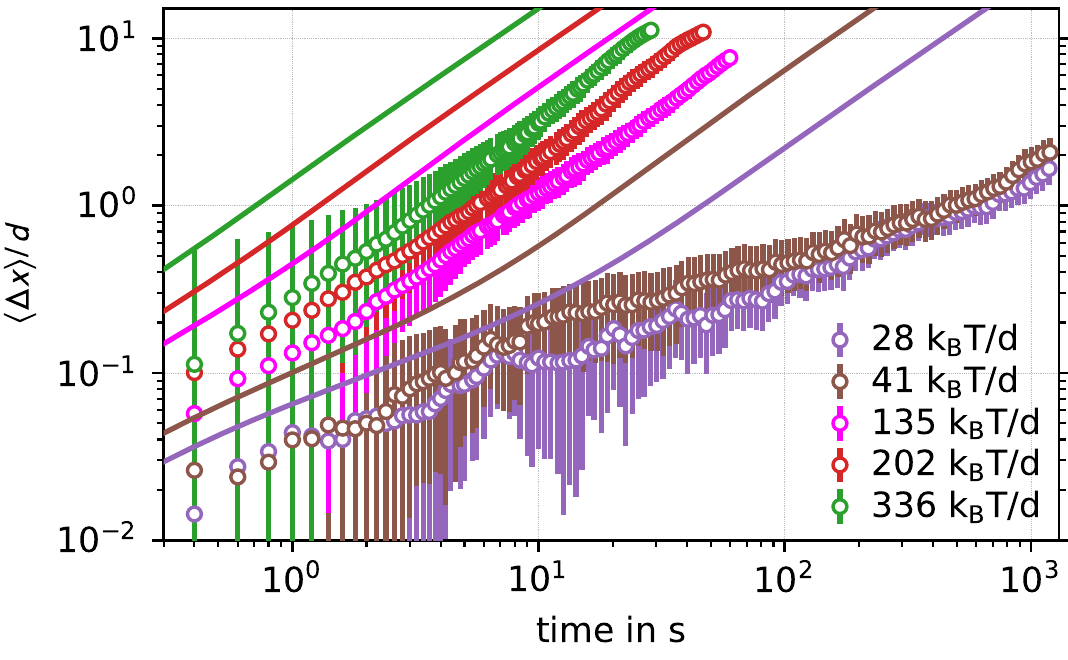}
 \includegraphics[width=.95\columnwidth]{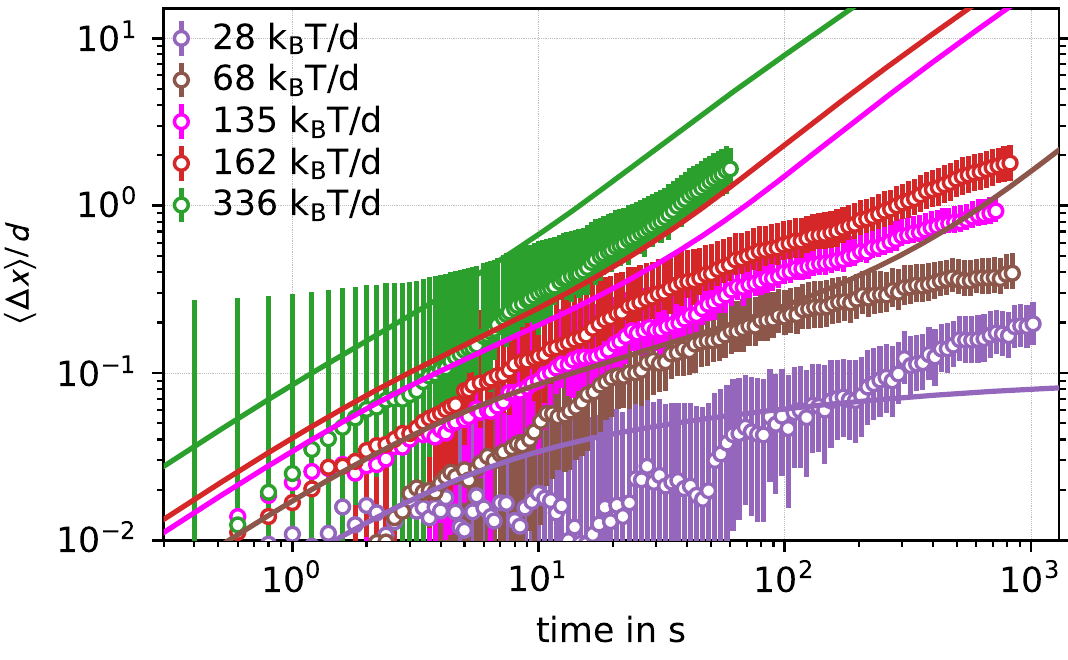}\\
 \caption{Mean displacements $\langle \Delta x(t) \rangle$ in liquid at $\phi=0.535$ (left), and in the glass $\phi=0.601$ (right). Error bars indicate the standard deviation of the mean and the systematic uncertainties for the start of the trajectories. Theory curves (solid lines) are obtained for the same force at the corresponding MCT packing fraction. No fitting is done. 
 }\label{fig_md}
\end{figure*}

\smallskip
\indent {\bf{Mean displacements.--}}
The raw mean displacements as obtained by averaging all particle positions at a given time are shown in Fig.~\ref{fig_md} in the liquid ($\phi=0.535$, left panel) and in the glass ($\phi=0.601$, right panel). Errorbars indicate statistical and systematic uncertainties; see the main text for discussion. The theory curves are obtained using the same forces as the experiments and the corresponding MCT packing fraction. In particular, no fitting has been done. We find a stronger localization in the experiment than predicted. We tentatively attribute a part of this difference to the use of a line-trap with a confining lateral potential. A build-up of bath particles in front of the probe may slow it down compared to when the perpendicular motion can fluctuate freely. 

\begin{figure}[ht]
 \includegraphics[width=.95\columnwidth]
{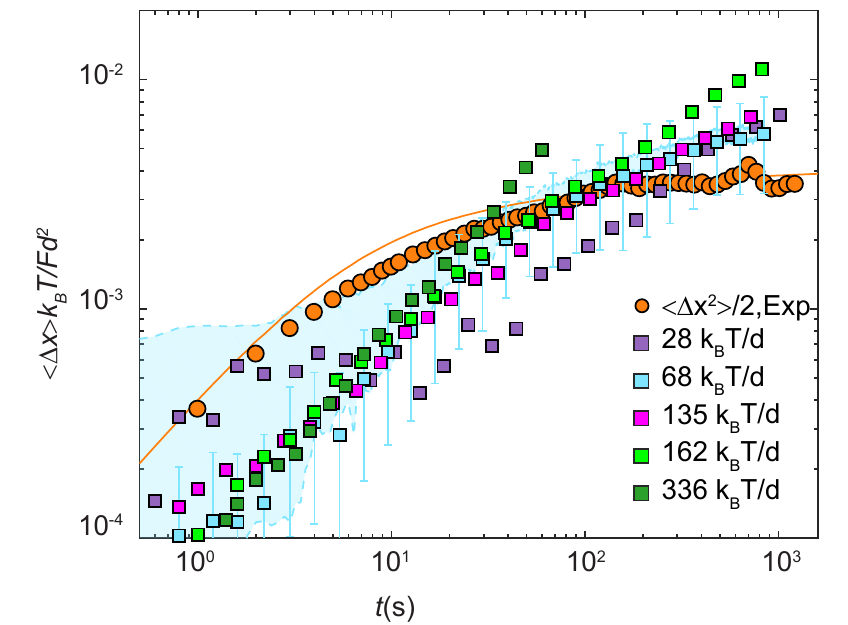}\\
 \caption{Rescaled mean displacements  in the glass at $\phi=0.601$; the corresponding plot for the fluid is shown as Fig.~2 in the main text. The raw displacements are divided by the non-dimensionalized force and compared to the quiescent mean squared displacement $\langle \Delta x^2\rangle_\text{eq}$ (circles). The shaded area indicates the uncertainty which arises from the difficulties in determining the starting position and time. Error bars (statistical uncertainties) are plotted for one value of the force only for clarity (full set is shown in Fig. \ref{fig_md}). 
}\label{figS6}
\end{figure}

\begin{figure*}[ht]
 \includegraphics[width=.95\columnwidth]{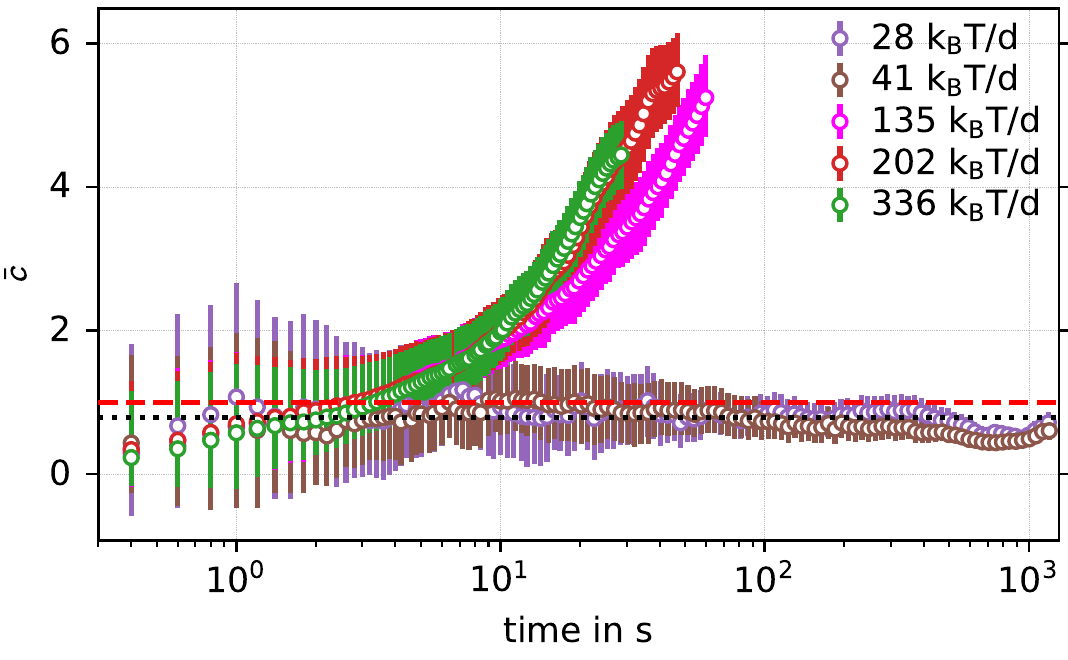}
 \includegraphics[width=.95\columnwidth]{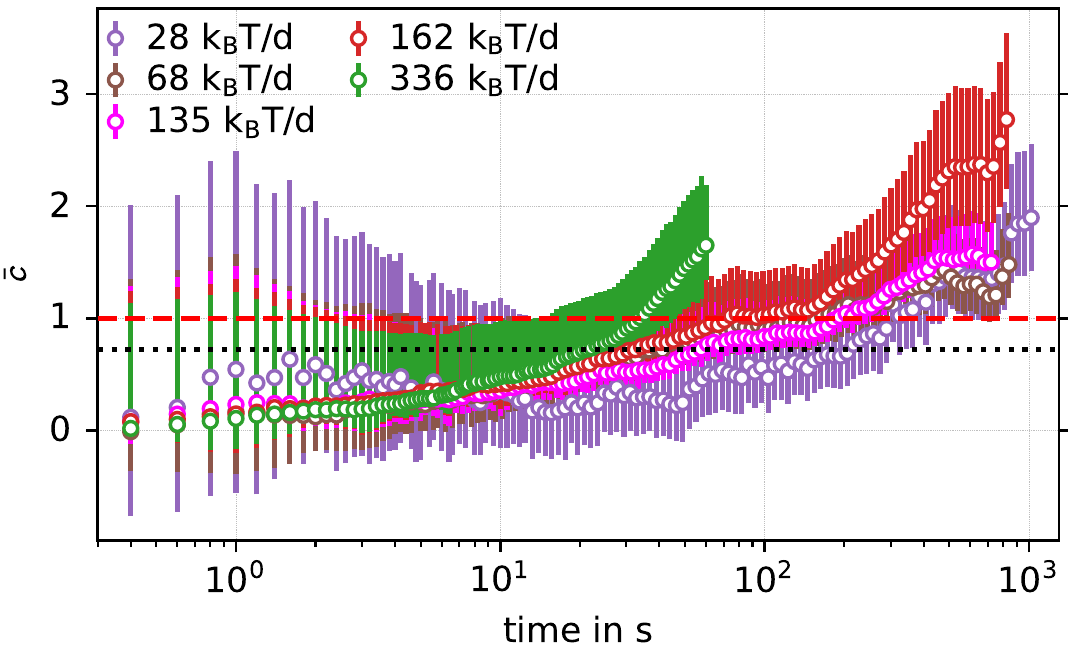}\\
 \caption{Linear response factor $\bar{c}$ (see Eq.~\eqref{Plscaling} for $\phi=0.535$ (left) and $\phi=0.601$ (right). The red dashed line shows the expected value $\bar{c}=1$. The dotted black line shows the value obtained by averaging all data-points for all forces for which linear response holds. 
 }\label{fig_lr}
\end{figure*}

\smallskip
\indent {\bf{Linear response calibration.--}}
In the laser tweezer calibration in a simple liquid it was found that the force $F$ of the laser tweezer is proportional to the actual laser power $P_l$ via $F=c P_l$ with $c=6.3$ $k_B T$/mW$=12.8$ fN/mW, Figure \ref{ltweezcal}. Although there is no reason for this proportionality to break down in the emulsion, there is the possibility of additional losses, e.g.\ due to scattering and optical abberations, which will reduce the prefactor of this proportionality relation. Therefore, we measure this prefactor using a linear response analysis in the emulsion.  The linear response relation Eq.~(1) relates the mean displacement to the mean squared displacement without external force, see Fig.~\ref{fig1}. In Fig.~2 of the main text  we show this relation for the fluid and in Fig.~\ref{figS6} we show it for the glass. For a more quantitative comparison, we calculate
\begin{equation}\label{Plscaling}
 \bar{c}(t) = \frac{2 \langle x(t)\rangle}{\langle \Delta x^2 (t)\rangle_\text{eq} } \frac{k_B T}{F}
\end{equation}
for every $t$ and every force available, which is shown in Fig. \ref{fig_lr}. From Eq. (1) we expect $\bar{c}(t)=1$ in the linear response regime. 
It can be seen that this relation holds quite well over the full range of times $t$ explored for small forces 
for $\phi=0.535$.  For the sample in the glass $\phi=0.601$ this relation is true for times between 50 and 100 seconds. The short times might be off due to the uncertainties about the initial position and starting time. Averaging all data points for forces below 100 $k_B T/d$ (for $\phi=0.535$) and below 180 $k_B T/d$ (for $\phi=0.601$) leads to $\bar{c}\approx 0.75$, indicating that the forces are about 25\% smaller than in the dilute system.

\smallskip
\indent {\bf{Theory details.--}}
The MCT approach to microrheology was developed in \cite{gruber2016active} and the calculation is performed using the algorithm described there. The time grid for the integrodifferential equations is uniform with 1024 grid points and an initial time step of $\Delta t = 10^{-8}$. Whenever the end of this grid is reached, the time step is doubled and the results are decimated. The $q$-space is discretized on a combined logarithmic and uniform grid. The uniform part of the grid has 101 points ranging from $qd=0$ to $qd=15$; in the logarithmic part the stepsize of the uniform grid is halved 10 times. This results in the following grid points: [0,1.5E-4,3E-4,\ldots,0.075,0.15,0.3,0.45,\ldots,15].
For this discretization of the system, the critical force is  $F_c = 34.3856869 k_B T / d$.  
\smallskip

\end{document}